\documentclass[aoas,MSNbibl,nameyear,dvips]{arximspdf}
\usepackage{dcolumn}
\usepackage{stfloats}
\usepackage{graphicx}

\doi{10.1214/14-AOAS784} % kopijuoti is laisko
\volume{8}
\issue{4}
\pubyear{2014}
\firstpage{2485}
\lastpage{2508}
\docsubty{FLA}

\makeatletter
  \fnbelowfloat
\def\mid{|}
\newcolumntype{d}[1]{D{.}{.}{#1}}
\newcommand{\rrvert}{\vert}
\newcommand{\rrVert}{\Vert}
\newcommand{\llvert}{\vert}
\newcommand{\llVert}{\Vert}
\newcommand{\eqref}[1]{(\ref{#1})}
\newcommand{\one}{\mathbf{1}}
\newcommand{\bE}{\mathbb E} % expectation

\newproclaim{ass}{Assumption}

\newcommand{\bD}{\mathbf{D}}
\newcommand{\bX}{\mathbf{X}}
\newcommand{\bV}{\mathbf{V}}
\newcommand{\bbeta}{\bolds{\beta}}
\newcommand{\blambda}{\bolds{\lambda}}
\newcommand{\PATT}{\mathrm{PATT}}

\makeatother

\begin{document}
\begin{frontmatter}

\title{Do debit cards increase household spending? Evidence from a
semiparametric causal analysis~of~a~survey}
\runtitle{A semiparametric causal analysis of debit cards usage}

\begin{aug}
\author[A]{\fnms{Andrea}~\snm{Mercatanti}\corref{}\ead[label=e1]{mercatan@libero.it}\thanksref{M1}}
\and
\author[B]{\fnms{Fan}~\snm{Li}\ead[label=e2]{fli@stat.duke.edu}\thanksref{M2}}
\runauthor{A. Mercatanti and F. Li}
\affiliation{Bank of Italy and Duke University}
\address[A]{Statistics Department\\
Bank of Italy\\
Via Nazionale, 91\\
00184 Rome\\
Italy\\
\printead{e1}}

\address[B]{Department of Statistical Science\\
Duke University\\
Durham, North Carolina 27708-0251\\
USA\\
\printead{e2}}

\end{aug}

% HISTORY:
\received{\smonth{2} \syear{2014}}
\revised{\smonth{8} \syear{2014}}

% ABSTRACT
%
\begin{abstract}
Motivated by recent findings in the field of consumer science, this
paper evaluates the causal effect of debit cards on household
consumption using population-based data from the Italy Survey on
Household Income and Wealth (SHIW). Within the Rubin Causal Model, we
focus on the estimand of population average treatment effect for the
treated (PATT). We consider three existing estimators, based on
regression, mixed matching and regression, propensity score weighting,
and propose a new doubly-robust estimator. Semiparametric specification
based on power series for the potential outcomes and the propensity
score is adopted. Cross-validation is used to select the order of the
power series. We conduct a simulation study to compare the performance
of the estimators. The key assumptions, overlap and unconfoundedness,
are systematically assessed and validated in the application. Our
empirical results suggest statistically significant positive effects of
debit cards on the monthly household spending in Italy.
\end{abstract}

% KEYWORDS
\thankstext{M1}{Supported in part by the U.S. National Science
Foundation (NSF)
under Grant DMS-11-27914 to the Statistical and Applied Mathematical
Sciences Institute (SAMSI).}
\thankstext{M2}{Supported in part by  NSF Grant SES-1155697.}
\end{frontmatter}

%s1 #&#
\section{Introduction}\label{sec1}
The past few decades have seen a steadily increasing global trend in
the use of noncash payment instruments such as credit, debit, charge
and prepaid cards as well as electronic money. Research on the
psychology of consumer behavior provides a theoretical basis for
supporting the thesis that payment instruments can play a significant
role in consumer decisions. Possibly the most important concept coming
out of this field of research is \textit{mental accounting}, that is,
the set of cognitive operations used by individuals and households to
organize, evaluate, and keep track of financial activities [Thaler
(\citeyear{Tha85,Tha99})]. Starting from this concept, recent research has proposed
theories on prospective accounting, coupling, retrospective
evaluations, and financial resources accessibility [\citet{PreLoe98}, \citet{Som01}, \citet{MorHotEpl07}], which have stimulated deeper investigation on the effects of
noncash payment instruments on consumption. Indeed, there has been
substantial evidence that consumers who have cards overspend compared
to those who do not [e.g., \citet{Bur74}, \citet{Hir79}, \citet{Tok93}, \citet{Col98}, \citet{Man06}]. However, the observed association
between the level of spending and the possession of cards does not
necessarily indicate the existence of causal links; the association
could be due to differences between the characteristics of card owners
and nonowners, or to differences in the situations in which cash and
cards are the preferred methods of payment. Despite the practical
importance of the problem and the large literature on causal inference
in statistics and economics, to our knowledge, there is essentially no
analysis on the causal effects of payment instruments on consumer spending.

The main objective of this paper is to investigate the causal effects
of debit cards possession on spending, using data from the Italy Survey
on Household Income and Wealth (SHIW) within the Rubin Causal Model
[RCM; Rubin (\citeyear{Rub74,Rub78}), \citet{Hol86}]. Debit cards are defined as
cards enabling the holder to have purchases directly charged to funds
on his account at a deposit-banking institution [C.P.S.S. (\citeyear{ComSys01})], and
in Italy they are usually called ``\textit{carte Bancomat}.'' Our focus
on debit cards is principally motivated by the fact this payment
instrument does not allow consumers to incorporate additional long-term
sources of funds, as in the case of credit cards. By considering debit
cards, it is therefore possible to eliminate the confounding
intertemporal reallocations of wealth from the psychological effects on
spending [\citet{Som01}]. Alternatively, prepaid cards could be relevant
to the current study's objectives, because they do not allow the
consumer to be granted a line of credit. However, their diffusion in
Italy is at the moment low.

Under the RCM, each unit has a potential outcome corresponding to each
treatment level, and the causal effect is defined as a comparison
between the potential outcomes of a common set of units. Ideally we
would conduct an analysis with units being individuals possessing debit
cards, because debit cards are typically issued to individuals.
However, SHIW collects information only on the household level. To
mitigate this problem, in our study, we set household as the unit, but
limit the sample of treated units to the households possessing one and
only one debit card. Such a choice ensures that a possible effect on
household spending will be due only to the individual possessing the
card, which is usually the head of the household. Formally, the
unit-level causal effect of holding debit card is defined as the
difference between the potential spending corresponding to with one and
only one debit card and without debit cards. In particular, we are
interested in the ``population average treatment effect for the
treated'' (PATT), that is, the average causal effect for the households
holding one debit card. The PATT identifies the change in the average
consumption for the households holding one debit card and due only to
the debit card, and thus provides a scientifically sound answer to the
question of whether debit cards encourage spending.

Because at most one potential outcome is observed for each unit,
unit-level causal effects are generally not identifiable. Nevertheless,
population average causal effects can be identified under some
assumptions. The most important and widely adopted identifying
assumption is unconfoundedness [\citet{RosRub83}], which
rules out self-selection into the treatment. Another key assumption is
overlap, which ensures overlap in covariate distributions between
groups. We maintain both assumptions throughout the paper. An integral
component of our application is to systematically assess, directly or
indirectly, the plausibility of these assumptions.

We estimate the PATT from the SHIW data using three existing
estimators, based on regression, propensity score weighting, mixed
matching and regression, as well as a new doubly-robust (mixed
weighting and regression) estimator. To flexibly model the large number
of covariates, we choose to proceed from a semiparametric perspective
based on power series specifications. Over the last decade, non- and
semi-parametric methods have been revealed to be successful in
attaining desirable properties where standard parametric models fail.
In fact, under a semiparametric power series approach, the efficiency
bound for a causal effect estimator under unconfoundedness [\citet{Hah98}] is attained by a regression-based method [\citet{ImbNewRid05}] or by weighting on the estimated propensity scores
[\citet{HirImbRid03}]. Other advantages include the
following: first, a correction based on power series regressions allows
for a matching method to be unbiased and consistent [\citet{AbaImb06}]; second, the overlap assumption can be easily assessed by
relying on undersmoothed specification for the propensity score [\citet{Imb04}]; and, finally, the assessment of unconfoundedness can be
performed by testing equality restrictions on the coefficients of power
series regressions for the treated and untreated units [\citet{Cruetal08}]. Despite these advantages, power series-based semiparametric
approaches have not been widely used in practice.

The rest of the paper is organized as follows: Section~\ref
{sec:motivation} presents the theories and some empirical findings,
mainly from the consumer psychology literature, that motivate the
current research problem. Section~\ref{sec:methods} sets up the causal
approach, introduces the new doubly-robust estimator and three existing
estimators, and describes the semiparametric specification. A small
simulation study is carried out in Section~\ref{sec:simulations} to
compare the estimators. In Section~\ref{sec:application}, we first
present some preliminary results of the real data, then assess the key
assumptions, and finally apply the semiparametric methods to estimate
the causal effects of possessing debit cards on household spending.
Section~\ref{sec:conclusion} concludes.

%s2 #&#
\section{Motivating background} \label{sec:motivation}
Research in psychology and consumer science shows that consumers are
highly sensitive to contextual information that may induce perceptual
contrasts when making evaluations [\citet{Hel64}, \citet{Hseetal99},
\citet{KahMil86}, Morewedge, Hotzman and
Epley (\citeyear{MorHotEpl07}), \citet{Par95}].
This field of research provides arguments for supporting that a
consumer's evaluation of the amount of disposable financial resources
can be heavily influenced by the size of the financial resources
cognitively, or temporarily, accessible at the time of purchasing [\citet{HeaSol96}, \citet{SomChe02}, \citet{Tha85}]. Based on
results from one small experiment, Morewedge, Hotzman and
Epley (\citeyear{MorHotEpl07}) suggested
that consumers perceive a unit of consumption to be cheaper when large,
as opposed to, small financial resources are made cognitively
accessible. As a result, large financial accounts, such as the money in
one's savings account, are likely to increase the likelihood of
consumption as compared to small financial resources, such as the
amount of money in one's wallet. But Morewedge et al. did not explain
when consumers are likely to think in terms of small versus large
disposable resources. It is reasonable to postulate that the method of
payment can activate thoughts about different financial resources.
Therefore, it would be useful to investigate the effects of payment
instruments that provide direct access to larger financial resources,
such as debit or credit cards, on consumption.

A second motivation stems from the findings of \citet{Som01}, who showed
that payment instruments influence the memory for and the impact of
past expenses on spending behavior. Two features of payment
mechanisms---rehearsal of the price paid and the immediacy of wealth
depletion---were manipulated using a controlled experiment in which
recall and retrospective evaluation of payments were measured
simultaneously with the purchase intention. The experiment involved
four different payment instruments: checks, debit cards, charge cards,
and charge checks. Debit cards are here characterized by no rehearsal
(like charge cards) in that consumers do not need to write down the
total amount; rather, they involve immediate wealth depletion (like
checks). Soman's study shows that past payments significantly reduce
future spending intentions when the payment instrument requires
consumers to write down the amount paid as well as when the consumer's
wealth is depleted immediately rather than with a delay.

Some attempts to quantify the effects of payment instruments,
especially credit cards, on spending were conducted by small-scale
randomized experiments [\citet{Fei86}, \citet{PreSim01},
\citet{Som01}]. All of these studies were performed on a small sample of
volunteers, raising the concern of external validity, as there may be
significant difference between the volunteers and the targeted
population. Moreover, all but one of the experiments involve only
simulated series of payments rather than real monetary transactions.
Also, the experiment in \citet{Fei86} that is based on real monetary
transactions only manipulates exposure to credit card stimuli, not the
payment method itself.

Population-based observational studies generally do not match the
internal validity of randomized experiments, but they usually offer
better external validity. Therefore, a careful causal analysis on a
large population-based observational data with information on real
monetary transactions, which was absent in the literature to our
knowledge, would provide a good complement to these randomized studies.
This motivates our analysis of the SHIW data, a biennial, nationally
representative survey run by Bank of Italy aiming to collect
information on several aspects of Italian households' economic and
financial behavior. SHIW contains rich information related to household
characteristics, spending, and payment instruments, and thus provides a
great opportunity to evaluate the causal effect of debit card
possession on spending in Italian households.

%s3 #&#
\section{Causal inference} \label{sec:methods}
%s3.1 #&#
\subsection{Setup, estimand, and assumptions}
Consider a large population of units, each of which can potentially be
assigned a treatment indicated by $z$, with $z=1$ for active treatment
and $z=0$ for control. A random sample of $N$ units from this
population is drawn to evaluate the treatment effect on some outcome.
For each unit~$i$ $(i=1,\ldots,N)$, let $Z_i$ be the observed treatment
status, and $\bX_i=\{X_{i1},\ldots,X_{ir}\}$ be a set of $r$ pre-treatment
variables (i.e., covariates) and the $N \times r$ matrix $\bX$ has
$i$th row equal to $\bX_i$. We assume the Stable Unit Treatment Value
Assumption [SUTVA; Rubin (\citeyear{Rub80})], that is, no interference between the
units and no different versions of a treatment. Then each unit $i$ has
two potential outcomes $Y_i(0)$ and $Y_i(1)$, corresponding to the
potential treatment levels $z=0$ and $z=1$, respectively. Between the
two potential outcomes, only the one corresponding to the observed
treatment status, $Y_i=Y_i(Z_i)$, is observed. In our study, the unit
is the household; the treatment status equals one if the household
possesses one and only one debit card and zero if the household does
not possess debit cards; and the outcome is the monthly household
spending on all consumer goods. SUTVA is deemed reasonable in this
setting, as the holding of debit cards in one household does not seem
to affect the potential spending of other households.

Our primary interest lies in the causal effect of having debit card on
spending for the households who possess one and only one debit card.
Therefore, the target causal estimand is the population average
treatment effect for the treated (PATT):
%
%e1 #&#
\begin{equation}
\label{eq:PATT} \tau^{\PATT}\equiv\bE\bigl\{Y_i(1)-Y_i(0)
\mid Z_i=1\bigr\}.
\end{equation}
To identify the PATT, we maintain the standard assumption of unconfoundedness.

%as1 #&#
\begin{ass}[(Unconfoundedness)]\label{ass1} The treatment assignment is independent
of the potential outcomes given a vector of pre-treatment covariates
$\mathbf{X}_i$:
\[
\bigl\{Y_{i}(0),Y_i(1)\bigr\} \perp Z_i
\mid\mathbf{X}_i.
\]
\end{ass}

Unconfoundedness assumes that the treatment assignment is randomized
conditional on a set of pre-treatment covariates, and thus rules out
self-selection into the treatment. It is also referred to as the
assumption of ``no unmeasured confounders.'' Under unconfoundedness, we
have $\Pr(Y_i(z)|\mathbf{X}_i)=\Pr(Y_i|Z_i=z, \mathbf{X}_i)$, and thus
causal effects can be estimated by the average difference in the
observed outcome between the groups that have balanced covariate
distributions. However, unconfoundedness, sometimes questionable in
observational studies, is generally untestable. Nevertheless, there are
indirect ways to assess its plausibility. In particular, we will adopt
the proposal of \citet{Cruetal08} to assess unconfoundedness by a
test on a pseudo-outcome, namely, the lagged outcome in this application.

The second assumption ensures overlap in the covariate distributions
between the treatment and control groups.

%as2 #&#
\begin{ass} [(Overlap)]\label{ass2} Each unit in the population has a nonzero
probability of receiving each treatment:
\[
0<e(\mathbf{x})\equiv\Pr(Z_i=1|\mathbf{X}_i=
\mathbf{x})<1\qquad \forall \mathbf{x},
\]
where $e(\mathbf{x})$ is called the propensity score [\citet{RosRub83}]. Violation to the overlap assumption generally leads to
operational difficulties, such as large variances in weighting
estimators, as well as conceptual difficulties because the potential
outcome under one treatment level for certain values of covariates
would never be observed and the causal effect would be {a priori}
counterfactual. The overlap assumption is directly testable, for
example, by visually inspecting the distributions of the estimated
propensity scores between groups. The combination of Assumptions \ref{ass1} and
\ref{ass2} is referred to as ``strong ignorability'' [\citet{RosRub83}].
\end{ass}

When the interest is in estimating the PATT, the outcome distribution
for the treated is directly estimable so that the two assumptions can
be slightly weakened [Heckman, Ichimura and Todd (\citeyear{HecIchTod98})] and replaced
by unconfoundedness only for the untreated units, $Y_{i}(0) \perp Z_i
\mid\mathbf{X}_i$, and with the weak overlap, $\Pr(Z_i=1|\mathbf
{X}_i=\mathbf{x})<1$ for any $\mathbf{x}$.

%s3.2 #&#
\subsection{Estimators} \label{sec:estimators}
We first introduce three existing estimators for the\break PATT. Let $\mu
_z(\mathbf{x})=\bE\{Y_i(z)|\bX_i=\mathbf{x}\}$ be the regression
function for the
potential outcome $Y(z)$, for $z=0,1$. The first estimator is based on
the estimation of the regression function for the untreated units $\mu
_0(\mathbf{x})$, from which the counterfactual outcome for the treated unit
$i$ can be predicted as $\hat{\mu}_0(\bX_i)$. The estimated PATT is
obtained from averaging the observed and the predicted outcomes of the
treated, as dated back from the parametric version of this estimator
[\citet{Rub77}]:
%
%e2 #&#
\begin{equation}
\label{eq:est_reg} \hat{\tau}_{\mathrm{reg}}= \sum_{i=1}^N
Z_{i} \bigl\{Y_{i}-\hat{\mu}_0(\bX
_i) \bigr\} \bigg/{\sum_{i=1}^N
Z_{i}}.
\end{equation}
A parametric prediction of $\hat{\mu}_0(\mathbf{x})$, however, would be
sensitive to differences in the distributions of the pre-treatment
variables between the treatment groups, which would make the estimation
procedure rely heavily on the functional specification [\citet{Imb04}].
Alternatively, Imbens, Newey and
Ridder (\citeyear{ImbNewRid05}) showed that the estimator with a
nonparametric estimation of $\hat{\mu}_0(\mathbf{x}_i)$ based on
power series
can achieve the nonparametric efficiency bound for the PATT [\citet{Hah98}].

The second estimator is based on propensity score weighting, originated
from the inverse probability weighting technique in survey sampling
[\citet{HorTho52}]. It is easy to show that
\[
\bE \biggl\{Y_iZ_i-\frac{Y_i(1-Z_i)e(\bX_i)}{1-e(\bX_i)} \biggr\} =\tau
^{\PATT}. \label{wt}
\]
Therefore, one can define the ATT weights for each unit: $w_i=1$ for
the treated units ($Z_i=1)$ and $w_i=\hat{e}(\bX_i)/(1-\hat{e}(\bX_i))$
for the control units ($Z_i=0)$, where $\hat{e}(\bX_i)$ is the
estimated propensity score for unit $i$. The weighting estimator for
the PATT with the sum of the weights in each group being normalized to
one [\citet{HirImb01}] is
%
%e3 #&#
\begin{equation}
\label{eq:est_wt} \hat{\tau}_{\mathrm{wt}}= \frac{\sum_{i=1}^N Y_iZ_iw_i}{\sum_{i=1}^N Z_i w_i} -\frac{\sum_{i=1}^N Y_i(1-Z_i)w_i}{
\sum_{i=1}^N(1-Z_i)w_i}.
\end{equation}
Here the (estimated) propensity scores are used to create a weighted
sample of untreated units that has the same covariate distribution as
that in the treated group [\citet{LiMorZas14}]. Hirano, Imbens and
Ridder (\citeyear{HirImbRid03}) showed that the efficiency bound for PATT estimators can be
achieved by weighting on the power series logit estimates of the
propensity scores.

The third estimator is a mixed matching and regression approach. A
standard matching estimator for PATT is obtained as follows: first, for
each treated unit $i$, find $m$ closest matched untreated units
according to a metric defined in the space of the covariates; second,
take the average of the observed outcome of the $m$ matches as the
estimated counterfactual outcome $\hat{Y}_{i}(0)$; and finally estimate
the PATT by the average of the estimated individual effects of the
treated units. Matching estimators ensure good balance in covariates
between groups and are generally robust [see, e.g., \citet{IchMeaNan08}]. However, if the number of matches is fixed and
matching is done with replacement, \citet{AbaImb06} showed the
bias of this estimator is $O(N^{-{1}/{p}})$, where $p$ is the number of
continuous covariates, while the variance of the estimator is
$O(N^{-1})$. In our study, $p=6$ so that, asymptotically, the bias will
not disappear and the standard confidence interval will not be
necessarily valid. To improve the asymptotic properties of matching
estimators, \citet{AbaImb11} proposed a mixed method where for
each treated unit, matching is followed by local regression
adjustments, which adjust for the residual differences in the
covariates between the treated unit and its matches:
%
%e4 #&#
\begin{equation}
\label{eq:est_reg} \hat{\tau}_{\mathrm{mix}}={\sum_{i=1}^{N}Z_{i}
\biggl[ Y_{i}-\sum_{j\in
\mathcal
{M}_{i}} \bigl\{
Y_{j}+\hat{\mu}_{0}(\bX_{i})-\hat{
\mu}_{0}(\bX _{j}) \bigr\}\Big/m \biggr]} \bigg/{\sum
_{i=1}^{N}Z_{i}},
\end{equation}
where $\mathcal{M}_i$ is the set of the indices of the $m$ closest
matches of unit $i$, and $\hat{\mu}_{0}(\mathbf{x})$ is the
predicted outcome
from a regression estimated using only the matched sample. If $\hat
{\mu
}_{0}(\mathbf{x})$ is estimated from a power series regression, the resulting
PATT estimator can be proven to be consistent and asymptotically
normal, with its bias dominated by the variance.

Finally, we propose a new mixed estimator for the PATT that combines
weighting and regression:
%
%e5 #&#
\begin{equation}\quad
\label{eq:est_dr} \hat{\tau}_{\mathrm{dr}}=\frac{\sum_{i=1}^N Y_i Z_i}{\sum_{i=1}^N Z_i} - \frac
{1}{\sum_{i=1}^N Z_i}
\sum_{i=1}^N\frac{Y_i(1-Z_i)\hat{e}(\bX
_i)+\hat{\mu
}_0(\bX_i)(Z_i-\hat{e}(\bX_i))}{1-\hat{e}(\bX_i)}.
\end{equation}
This estimator requires specifying models for both potential outcomes
and propensity score. We can prove $\hat{\tau}_{\mathrm{dr}}$ is
``doubly-robust'' (DR) (see the\break \hyperref[app]{Appendix}), that is, it has the large
sample property that the estimator is consistent if either the
propensity score model or the potential outcome model is correctly
specified, but not necessarily both [\citet{RobRotZha95}]. The existing literature on DR estimators has exclusively
focused on the average treatment effect (ATE) estimand. To our
knowledge, estimator \eqref{eq:est_dr} provides the first explicit DR
estimator for the PATT. Like the DR estimator of the PATE, \eqref
{eq:est_dr} is a member of the class of consistent, efficient,
semiparametric estimators of Robins, Rotnitzky and
Zhao, where the numerator of the
second term has the form of that in a weighting estimator but augmented
by an expression involving the regression for the outcomes.

Besides the main theoretical advantage of robustness compared to the
weighting or the regression estimator, the DR estimator can also serve
as a diagnostic tool in practice: if the DR estimate differs much from
the regression estimate, but is similar to the weighting estimate, it
would suggest a potential misspecification of the regression model
(e.g., an incorrect choice of the order term of the power series or
lack of interaction term) or a lack of overlap. Alternatively, if the
DR estimate differs from the weighting estimate, but is similar to the
regression estimate, it would suggest a potential misspecification of
the propensity score model, which is possible even if the visual check
of the estimated propensity scores suggests sufficient overlap.

Both $\hat{\tau}_{\mathrm{mix}}$ and $\hat{\tau}_{\mathrm{dr}}$ are mixed approaches:
combining regression with matching or weighting. Weighting and matching
have distinct operating characteristics: weighting is a ``top-down''
approach in the sense that it applies weights to the entire sample and
is designed to create global balance for the target population, whereas
matching is a ``bottom-up'' approach in the sense that it limits the
analysis to the matched subsample and is designed to create local
balance for this subsample [Li, Morgan and Zaslavsky (\citeyear{LiMorZas14})]. Both methods have pros
and cons, and there is no universal rule for choosing between them,
which highly depends on the goal and practical constraints of a
specific study. As shown in the simulations, the mixed-matching
estimator is more robust than the DR estimator, but is less efficient
when there is no misspecification.

Different ways to calculate the standard errors have been adopted for
these PATT estimators. The delta method and the bootstrap lead to valid
standard errors when the estimators are based on series estimates of
the regressions and/or the propensity scores. Here we adopt bootstrap
for $\hat{\tau}_{\mathrm{reg}}$, $\hat{\tau}_{\mathrm{wt}}$, and $\hat{\tau}_{\mathrm{dr}}$.
Bootstrap is not valid for matching methods with a fixed number of
matches, and the standard errors for $\hat{\tau}_{\mathrm{mix}}$ have been
obtained using the estimator proposed by \citet{AbaImb06}.

%s3.3 #&#
\subsection{Semiparametric specification} \label{sec:semi_spec}
All of the four estimators require specifying regression functions for
either potential outcome $\mu_z(\mathbf{x})$ or propensity score
$e(\mathbf{x})$ or
both. Parametric specification is the standard approach in the
literature. However, parametric methods are usually sensitive to
imbalance between groups and misspecification, a serious concern in
observational studies with a large number of covariates. Nonparametric
specification is flexible and less prone to misspecification, but is
often difficult to estimate due to the potentially large number of
parameters. Therefore, in this paper we choose the semiparametric
specification based on power series [\citet{HauNew95}, \citet{DasNewVel03}] for both potential outcome and propensity
score, which combine the virtues and mitigate the problems of
parametric and nonparametric approaches.

The main idea is to divide the covariates $\bX$ into two groups $\bX
_{(1)}$ and $\bX_{(2)}$, and specify a semiparametric, partially linear
model for the mean function:
%
%e6 #&#
\begin{equation}
\label{eq:semi_general} \mu (\mathbf{X}; \bbeta ) =\bX_{(1)}\bbeta+g(
\bX_{(2)}),
\end{equation}
where $\bX_{(1)}$, with dimension $h$, enters the model in a simple
linear fashion (as main effects), and $\bX_{(2)}$, with dimension $s$,
enters the nonparametric part of the model, $g(\bolds{\cdot})$. We now
give the general (and somewhat complex) form of power series
specification of $\mu(\bolds{\cdot})$, followed by a simple example
used in our application. Let $r=h+s$ be the dimension of the argument
of $\mu(\bolds{\cdot})$, and $\blambda=(\lambda_{1},\ldots
,\lambda
_{r})$ be a multi-index of nonnegative integers with norm $\llvert
\blambda\rrvert =\sum_{j=1}^{r}\lambda_{j}$. Let $\bX
_{i}^{\blambda}=X_{i1}^{\lambda_{1}}\cdots X_{ir}^{\lambda
_{r}}$ be the product of the powers of the components of $\bX_{i}$, and
$ \{ \lambda(k) \} _{k=1}^{\infty}$ a series of distinct
multi-index such that $\llvert \lambda(k)\rrvert $ is
nondecreasing. Let $p_{k}(\mathbf{X}_{i})=\bX_{i}^{\lambda(k)}$,
$\mathbf{p}_{k}(\mathbf{X}_{i})=(p_{1}(\mathbf{X}_{i}),\ldots
,p_{k}(\mathbf
{X}_{i}))'$, and finally $\mathbf{P}_{k}(\bX)=(\mathbf
{p}_{k}(\mathbf
{X}_{1}),\ldots,\mathbf{p}_{k}(\mathbf{X}_{N}))'$. Given the particular
order term $k$, the series estimator of the regression function $\mu
 ( \mathbf{X}; \bbeta )$ under the treatment $z$ ($z=0,1$) is
%
%e7 #&#
\begin{equation}
\label{eq:beta_est} \hat{\mu}_{z} ( \mathbf{X}; \bbeta ) =
\bD_{k,z}\bigl(\bD _{k,z}'\bD _{k,z}
\bigr)^{-1}\bD_{k,z}'\tilde{\mathbf{Y}}_z,
\end{equation}
where the design matrix is $\bD_{k,z}=\mathbf{P}_k(\tilde{\bX}_{z})$,
with $\tilde{\bX}_{z}, \tilde{\mathbf{Y}}_z$ being the matrices and the
vector of all the observed values of $\bX, \mathbf{Y}$ in group $z$.

In our application, we choose $\bX_{(1)}$ to contain all the dummy
variables of discrete covariates, $\bX_{(2)}$ to contain all the
continuous covariates, and $g(\bolds{\cdot})$ to be a polynomial for each
variable in $\bX_{(2)}$ with the same maximum power term $l$ excluding
interactions. Here $k=1+h+ s \times l$. A simple example is when there
is only one discrete variable, $\bX_{i(1)}=\{X_{i1}\}$, and one
continuous variable, $\bX_{i(2)}=\{X_{i2}\}$. Let $\blambda=(\lambda
_{1},\lambda_{2})$, $\mathbf{X}_{i}^{\lambda}=X_{i1}^{\lambda
_{1}}\cdot X_{i2}^{\lambda_{2}}$, and the sequence of $\lambda(k)$
with nondecreasing norm be
\begin{eqnarray*}
\lambda(1)&=&(0,0), \qquad
\lambda(2)=(1,0), \qquad
\lambda(3)=(0,1),\\
 \lambda (4)&=&(0,2),\qquad \ldots,\qquad
\lambda(k)=(0,l).
\end{eqnarray*}
Then the generic term $\mathbf{p}_{k}(\mathbf
{X}_{i})=(1,X_{i1},X_{i2},X_{i2}^{2},\ldots,X_{i2}^{l})'$. If $l=3$, then
$k=5$ and the mean model is
\[
\mu ( \mathbf{X}_i; \bbeta ) = \beta_0 +
X_{i1}\beta_1 +X_{i2}\beta_{21}+X_{i2}^2
\beta_{22}+X_{i2}^3\beta_{23}.
\]

A key component in the implementation is the choice of the order term
$k$. We will adopt the standard ``leaves-one-out'' cross-validation (CV)
approach, which selects the $k$ that minimizes the mean squared errors
(MSE) when predicting the outcome of each unit from all other units.

For the propensity score, we assume
%
%e8 #&#
\begin{equation}
\label{eq:ps} \log \biggl\{\frac{e(\bX)}{1-e(\bX)} \biggr\} = \mu_e(\bX;
\bbeta_e),
\end{equation}
where $\mu_e(\bX; \bbeta_e)$ is specified as in model \eqref
{eq:semi_general}. The closed-form least square estimator for $\mu_e$
is generally unavailable and the parameters are estimated via numerical
methods. The order term of the power series is also chosen from CV
based on minimizing a MSE criterion, where the predicted error for each
unit is the difference between the observed $Z_i$ and the estimated
propensity score $\hat{e}(\bX_i)$. Such a choice is driven by the
balance between the bias and the variance of $\hat{e}(\bX)$. In
estimating propensity scores, we are mainly interested in reducing the
bias rather than the variance; particularly we want to obtain
propensity score estimates that balance the covariates between the
groups. For this reason, \citet{Imb04} recommends to adopt higher order
power series than the one chosen by CV, that is, undersmooth the
estimation of the propensity score, and thus reduce the risk of failure
in detecting a lack of overlap in covariate distributions.

%s4 #&#
\section{Simulations} \label{sec:simulations}
\begin{figure}

\includegraphics{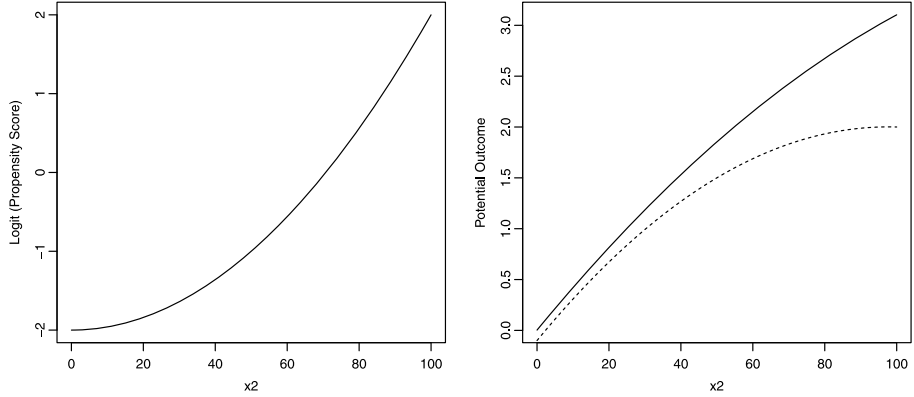}

\caption{Logit of the propensity score (left) and the potential
outcomes [right, solid line $\bE\{Y(1)\}$, dashed line $\bE\{Y(0)\}$]
as a function of $X_2$ with $X_1=0$ in the simulations.} \label{fig:simulation}\vspace*{-3pt}
\end{figure}

To compare the performance of the four estimators, we conduct a small
simulation study. The hypothetical population is set to be two groups
of units with different distributions of pre-treatment covariates; each
unit has a continuous outcome $Y$ and two covariates, a binary $X_{1}$
and a continuous $X_{2}$. This is to mimic a real situation where, for
example, the population consists of two groups with the minority being
a group of people with higher social-economic status; $Y$ is the
consumption, $Z$ is a payment instrument, for example, debit card, and $Z=1$
means possessing a card, and $X_1$ and $X_{2}$ are education and
income, respectively. The variables $X_1$ and $X_2$ are drawn from the
following distributions:
\begin{eqnarray*}
X_1&\sim&\mbox{Bernoulli}(0.25),\\  X_2 \mid
X_1 &\sim&\one\{X_{1}=1\} \cdot N(80,20)+\one
\{X_{1}=0\}\cdot N(20,20).
\end{eqnarray*}
Both the true propensity score and the potential outcomes models are
set to be nonlinear (quadratic) functions of $X_{2}$, and the
parameters are chosen based on the estimated coefficients of education
and income in the corresponding model from the real data. Specifically,
the true propensity score model, shown in Figure~\ref{fig:simulation},
left panel (with $X_1=0$), is
\[
\operatorname{logit}\bigl\{e(\bX)\bigr\}=-2+X_{1}+0.0004X_{2}^{2},
\]
so that the propensity score is equal to $0.14$ if $X_{1}=0$ and
$X_{2}=20$, to $0.83$ if $X_{1}=1$ and $X_{2}=80$. The true potential
outcomes models, shown in Figure~\ref{fig:simulation}, right panel
(with $X_1=0$), are
\begin{eqnarray*}
Y(0)&=&-0.1+0.1X_{1}+0.043X_{2}-0.00022X_{2}^{2}+
\varepsilon,
\\
Y(1)&=&0.1X_{1}+0.043X_{2}-0.00012X_{2}^{2}+
\varepsilon,
\end{eqnarray*}
where $\varepsilon\sim N(0,0.15)$. The figures show that the first
derivative of the propensity score increases with $X_{2}$ while the
first derivative of the potential outcomes decreases with $X_{2}$.
These models lead to a true PATT of $0.5778$ (evaluated on large samples).

%f1 #&#

We generate 500 simulated samples, each consisting of 1000 units. For
each unit, we first generate $X_1$ and $X_2$, and then generate two
potential outcomes, and the propensity score, based on which treatment
status $Z$ is drawn. For each simulated sample, we apply the
semiparametric model selected by CV in comparison to the simple linear
specification models $\beta_0 +\beta_{1}X_{1}+\beta_{2}X_{2}$, for each
estimator. The DR estimator has been evaluated also for the two cases
where the potential outcomes models or the propensity score model are
set to be linear (misspecified). The number of matches in the
mixed-matching estimator is set to 6.

%t1 #&#
\begin{table}[b]\vspace*{-3pt}
\caption{Average absolute bias (bias), root mean square error (RMSE),
and coverage of the 95\% confidence interval (coverage) of different
estimators in the simulation. The estimators $\hat{\tau}_{\mathrm{dr}}$ (p.s.)
and $\hat{\tau}_{\mathrm{dr}}$ (p.o.) represent the (partially misspecified) DR
estimator with the propensity score model and the potential outcome
model misspecified, respectively. Note that these two estimators,
though presented under the ``semiparametric'' category, are in fact
mixed-semiparametric--linear ones} \label{tab:simulation}
\begin{tabular*}{\textwidth}{@{\extracolsep{\fill}}lcccccc@{}}
\hline
& \multicolumn{3}{c}{\textbf{Semiparametric}} &
\multicolumn{3}{c@{}}{\textbf{Linear}}\\[-4pt]
& \multicolumn{3}{c}{\hrulefill} &  \multicolumn{3}{c@{}}{\hrulefill}\\
& \textbf{Bias} & \textbf{RMSE}&\textbf{Coverage} &
\textbf{Bias} & \textbf{RMSE} &\multicolumn{1}{c@{}}{\textbf{Coverage}}\\
\hline
$\hat{\tau}_{\mathrm{reg}}$ & 0.022 &0.035 &0.889 & 0.200 & 0.202& 0\phantom{0000} \\
$\hat{\tau}_{\mathrm{wt}}$ & 0.001 &0.058 &0.920 & 0.067 & 0.077& 0.879 \\
$\hat{\tau}_{\mathrm{mix}}$ & 0.009 &0.035 &0.939 & 0.062 & 0.072& 0.697 \\
$\hat{\tau}_{\mathrm{dr}}$ & 0.003 &0.034 &0.924 & 0.144 & 0.150& 0.159 \\
$\hat{\tau}_{\mathrm{dr}}$ (p.s.) & 0.016 &0.033 &0.897 & -- & -- & -- \\
$\hat{\tau}_{\mathrm{dr}}$ (p.o.) & 0.008 &0.138 &0.740 & -- & -- & -- \\
\hline
\end{tabular*}
\end{table}

Table~\ref{tab:simulation} reports the absolute bias, root of mean
squared error (RMSE), and coverage of the 95\%\vadjust{\goodbreak} confidence interval of
each estimator. Unsurprisingly, the semiparametric specification
dominates its linear counterpart for each estimator. Within the
semiparametric specification, the weighting and the DR estimator have
the smallest biases (0.001 and 0.003, resp.), with DR having a
lower RMSE $(0.034)$. Coverage of the 95\% confidence intervals are
similar between these two estimators (0.920 for $\hat{\tau}_{\mathrm{wt}}$ and
0.924 for $\hat{\tau}_{\mathrm{dr}}$). The mixed-matching estimator gives a
larger bias (0.009) but similar RMSE and coverage to those of the DR
estimator. The regression estimator gives the biggest bias (0.022) and
the lowest coverage (0.889). When only the propensity score model is
misspecified, the DR estimator still outperforms both the weighting and
the regression estimator. But when both the models are misspecified DR
leads to much higher bias and RMSE than the misspecified weighting
estimator. The mixed-matching estimator appears to be the least
sensitive to misspecifications among the four estimators, especially
when there is significant covariate imbalance between treatment groups,
as in this simulation. More simulations (omitted here) show this
advantage diminishes with increasing covariate balance or decreasing
sample size. Finally, the results demonstrate the aforementioned
diagnostic potential of the DR estimator: when only the propensity
score model is misspecified, the bias from the DR estimator is closer
to that obtained from a correctly-specified regression estimator, while
when only the potential outcomes model is misspecified, the bias from
the DR estimator is closer to that obtained from a correctly-specified
weighting estimator.

%s5 #&#
\section{Application to the Italy SHIW data} \label{sec:application}
%s5.1 #&#
\subsection{Data and preliminary analysis}
The SHIW has been run every two years since 1965 with the only
exception being that the 1997 survey was delayed to 1998. We denote by
$t$ the generic survey year, and by $(t+1)$ the subsequent survey year.
We define the target population as the set of households having at
least one bank current account but no debit cards at $t$. The treatment
$Z$ is posed equal to~1 if, at $t+1$, the household (all members
combined) possesses one and only one debit card, equal to 0 if, at
$t+1$, the household do not possess debit cards. The households with
more than one debit cards are excluded from the sample; therefore, a
household for which $Z=1$ is characterized by having acquired their
first (and only) debit card during the span $t\rightarrow(t+1)$.
Though we do not have exact information of the ownership of the debit
cards, it is reasonable to assume that it is the head of the household
who possesses the card in most cases. The outcome on which to evaluate
the treatment effect is the monthly\vadjust{\goodbreak} average spending of the household
on all consumer goods\setcounter{footnote}{2}\footnote{For the outcome, the relevant question
asks to consider all spending, on both food and nonfood consumption,
and it excludes only purchases of precious objects, purchases of cars,
purchases of household appliances and furniture, maintenance payments,
extraordinary maintenance of the dwelling, rent for the dwelling,
mortgage payments, life insurance premiums, and contributions to
private pension funds.} and is observed at $t+1$. The pre-treatment
variables include the following: the lagged outcome, some background
demographic and social variables referred either to the household or to
the head householder, the number of banks, and the yearly based average
interest rate in the province where the household lives. The subset of
pre-treatment variables referred to the household includes the
following: the number of earners (four categories), average age of the
household (five categories), family size (five categories), the overall
household income and wealth, the Italian geographical macro-area where
the household lives (three categories), the number of inhabitants of
the town where the household lives (four categories), and the monthly
average cash inventory held by the household. Those related to the head
householder include age (five categories) and education (six
categories). All the information is drawn from responses to the SHIW
questionnaires with the exception of the number of banks and the
average interest rate that are available since 1993 from the Bank of
Italy Monetary Statistics. These two variables have been suggested by
\citet{AttGuiJap02}, who showed, in a noncausal
context and to different purposes, that interest rate and the number of
banks in the area where the household lives have a significant
contribution to the probability of acquiring a debit card in Italy.

%t2 #&#
\begin{table}[b]
\caption{Sample sizes and relative frequency of treated and untreated
units for each span}
\label{tab:samplesize}
\begin{tabular*}{\textwidth}{@{\extracolsep{\fill}}lccccc@{}}
\hline
 & \multicolumn{2}{c}{\textbf{Treated}} & \multicolumn
{2}{c}{\textbf{Untreated}} & \\[-4pt]
 & \multicolumn{2}{c}{\hrulefill} & \multicolumn
{2}{c}{\hrulefill} & \\
$\bolds{{t\rightarrow(t+1)}}$& {\textbf{Size}} &
{\textbf{Rel. freq.}} & { \textbf{Size}} & {\textbf{Rel. freq.}} & \multicolumn{1}{c@{}}{\textbf{Total}}\\
\hline
1993--1995 & {223} & {0.217} &{805} & {0.783} & {1028} \\
1995--1998 & {188} & {0.322} &{396} & {0.678} & \phantom{0}{584} \\
1998--2000 & {160} & {0.230} &{534} & {0.770} & \phantom{0}{694} \\
\hline
\end{tabular*}
\end{table}

The PATT is estimated by comparing the observed outcomes for the
treated units with their predicted counterfactual outcomes. As a
consequence, reliable inferences need sufficiently large samples of
untreated units where to predict the counterfactual outcomes. SHIW is a
repeated cross-section with a panel component, namely, only a part of
the sample comprises households that were interviewed in previous
surveys. Our analysis will focus on the households observed for two
consecutive surveys. Table~\ref{tab:samplesize} reports the samples
sizes for each span, $t\rightarrow(t+1)$, where $t=1993,95,98$,
distinctly by treated and untreated units. The relative frequency of
untreated units (the households not possessing debit cards) alongside
the total sample size has a considerable drop after 2000. Accordingly,
the analysis will be limited to the span until 1998--2000, the latest
presenting considerable share of both untreated units and total sample size.

As a preliminary step, we conduct a simple descriptive cross-sectional
analysis on the subsample of households observed in a single sweep of
the survey. The sample size, shown in the first row of Table~\ref
{tab:prelim}, is considerably larger than that of the corresponding
year in Table~\ref{tab:samplesize}. The average difference in monthly
average spending between households possessing one debit card and
households without a debit card is 324.9, 307.8, and 437.3 thousands of
Italian Liras (the Italian currency until 2002) in year 1995, 1998, and
2000, respectively. Though not sufficient to establish causal effects
of debit cards on spending, this shows that consumers in Italy who
possess debit cards spend more compared to those who do not. To explore
the characteristics of the households possessing debit cards, we fit a
logistic model to this subsample, where the log odds ratio of having
one debit card at a certain year is linearly regressed on a set of
background demographic and social variables observed at the same year.
The variables and their estimated coefficients are shown in Table~\ref
{tab:prelim}. We observe significant contributions for many of the
explanatory variables for each year. In particular, the probability of
observing a household that has one debit card increases with income,
the town size, education of the head householder, from the South to
North of Italy, while it decreases with the average age of the household.

%t3 #&#
\begin{table}
\caption{MLE of the coefficients ($p$-values in parenthesis) of the
logistic model with response variable being the indicator of a
household having one debit card}
\label{tab:prelim}
\begin{tabular*}{\textwidth}{@{\extracolsep{4in minus 4in}}ld{2.8}d{2.8}d{2.8}@{}}
\hline
& \multicolumn{1}{c}{\textbf{{1995}}} & \multicolumn{1}{c}{\textbf{{1998}}} & \multicolumn{1}{c@{}}{{\textbf{2000}}} \\
\hline
{Sample size} & \multicolumn{1}{c}{\textup{4636}} & \multicolumn{1}{c}{\textup{4010}} & \multicolumn{1}{c}{4528} \\
{Intercept} & -1.80\ (0.00) & -0.02\ (0.96) & 0.30\ (0.46) \\
{Income} &\multicolumn{1}{c}{$1.3\times 10^{-5}$ (0.00)} & \multicolumn{1}{c}{$3.9\times 10^{-6}$ (0.00)} &
\multicolumn{1}{c}{$7.9\times 10^{-6}$ (0.00)} \\
{Wealth} & \multicolumn{1}{c}{$-4.8\times 10^{-7}$ (0.00)} & \multicolumn{1}{c}{$-9.0\times 10^{-8}$ (0.19)} &
\multicolumn{1}{c@{}}{$-2.3\times 10^{-7}$ (0.00)} \\
\multicolumn{4}{@{}l}{Geographical area (baseline: North):} \\
{Center} & -0.46\ (0.00) & 0.01\ (0.92) & -0.15\ (0.18) \\
{South and Islands} & -0.77\ (0.00) & -0.36\ (0.00) & -0.60\ (0.00)
\\
\multicolumn{4}{@{}l}{Town size (baseline: $<$20,000):} \\
{20,000--40,000} & 0.21\ (0.02) & -0.04\ (0.64) & -0.11\ (0.25) \\
{40,000--500,000} & 0.40\ (0.00) & 0.13\ (0.12) & -0.02\ (0.75) \\
{$>$500,000 } & 0.60\ (0.00) & 0.28\ (0.10) & -0.11\ (0.48) \\
\multicolumn{4}{@{}l}{Family size (baseline: 1):} \\
{2} & 0.03\ (0.80) & 0.03\ (0.79) &  -0.01\ (0.88) \\
{3} & 0.08\ (0.59) & 0.18\ (0.24) & -0.17\ (0.23) \\
{4} & -0.00\ (0.98) & 0.03\ (0.84) & -0.09\ (0.58) \\
{$>$4} & -0.14\ (0.46) &-0.03\ (0.86) & -0.37\ (0.06) \\
\multicolumn{4}{@{}l}{No. of earners (baseline: 1):} \\
{2} & -0.09\ (0.25) & -0.04\ (0.62) & 0.11\ (0.18) \\
{3} & -0.05\ (0.69) & 0.12\ (0.41) & 0.09\ (0.50) \\
{$>$3} & -0.00\ (0.96) &-0.36\ (0.14) & -0.05\ (0.81) \\
\multicolumn{4}{@{}l}{Average age of the household (baseline: $<$31):} \\
{31--40} & -0.12\ (0.28) & -0.22\ (0.06) & -0.25\ (0.04) \\
{41--50} & -0.23\ (0.11) & -0.12\ (0.46) & -0.31\ (0.05) \\
{51--65} & -0.38\ (0.03) & -0.31\ (0.11) & -0.40\ (0.03) \\
{$>$65} & -1.09\ (0.00)& -0.90\ (0.00) & -1.31\ (0.00) \\
\multicolumn{4}{@{}l}{Education of the head of the household (baseline:
none):} \\
{Elementary school} & 0.31\ (0.12) & 0.14\ (0.52) & 0.70\ (0.00)
\\
{Middle school} & 0.94\ (0.00) & 0.68\ (0.00) & 1.06\ (0.00) \\
{Prof. 2nd school} & 0.98\ (0.00) &0.76\ (0.00) & 1.38\ (0.00) \\
{High school} & 1.32\ (0.00) & 1.09\ (0.00) & 1.64\ (0.00) \\
{University} & 1.36\ (0.00) & 1.19\ (0.00) & 1.76\ (0.00) \\
\multicolumn{4}{@{}l}{Age of the head of the household (baseline: $<$31):} \\
{31--40} & -0.05\ (0.73) & -0.07\ (0.70) & 0.22\ (0.24) \\
{41--50} & -0.25\ (0.15) & -0.33\ (0.11) & -0.10\ (0.60) \\
{51--65} & -0.40\ (0.04) & -0.41\ (0.07) & -0.23\ (0.28) \\
{$>$65} & -0.62\ (0.00)& -0.86\ (0.00) & -0.48\ (0.04) \\
{Average interest rate} & 0.21\ (0.05) & 0.07\ (0.68) & -0.30\ (0.15) \\
{Number of banks} & -0.00\ (0.16) & 0.00\ (0.13)& 0.00\ (0.31) \\
\hline
\end{tabular*}\vspace*{-3pt} 
\end{table}

%s5.2 #&#
\subsection{Model specification}
We estimate the propensity score and the potential outcomes according
to the semiparametric specification in Section~\ref{sec:semi_spec},
where the order term is selected from leave-one-out CV. For both
models, the covariates include those listed in Table~\ref{tab:prelim},
the cash inventory held by the household, and the lagged outcome from
the previous survey. Tables \ref{tab:CVp} and \ref{tab:CVmu0} report
the mean squared predicted errors (MSE) for the propensity score,
$e(\mathbf{x})$, and the outcome regression, $\mu_{z=0} (
\mathbf
{x} )$, respectively, where $l$ denotes the maximum power term in
the power series expansion of the nonparametric part $g(\bolds{\cdot
})$ in model \eqref{eq:semi_general}. The MSE for estimating the
propensity score is minimized, for each span, when the continuous
pre-treatment variables are posed at the simplest linear specification,
$l=1$. We then undersmooth the propensity score specification by
expanding $g(\bolds{\cdot})$ to $l=5$. For the outcome model,
according to Table~\ref{tab:CVmu0}, we set $l=1$ for the spans 1993--1995
and 1998--2000, and $l=3$ for the span 1995--1998.

%t4 #&#
\begin{table}
\tablewidth=200pt
\caption{MSEs for predicting propensity score $e(\mathbf{x})$, obtained
from leave-one-out cross-validation, and $^{\ast}$ denotes the
minimized MSE}
\label{tab:CVp}
\begin{tabular*}{200pt}{@{\extracolsep{\fill}}ld{1.5}d{1.5}d{1.5}@{}}
\hline
${\bolds{l}}$ & \multicolumn{1}{c}{\textbf{1993--1995}} & \multicolumn{1}{c}{\textbf{1995--1998}} & \multicolumn{1}{c@{}}{\textbf{1998--2000}}\\
\hline
{1} & 0.4076^{\ast} & 0.4748^{\ast} & 0.4082^{\ast}\\
{2} & 0.4099 & 0.4771& 0.4114 \\
{3} & 0.4097 & 0.4808 & 0.4201 \\
{4} & 0.4120 & 0.4862 & 0.4204 \\
{5} & 0.4126 & 0.4886 & 0.4265 \\
\hline
\end{tabular*}
\end{table}

%t5 #&#
\begin{table}
\tablewidth=200pt
\caption{MSEs for predicting ${\mu}_{z=0}(\mathbf{x})$, obtained from
leave-one-out cross-validation, and $^{\ast}$ denotes the minimized
MSE} \label{tab:CVmu0}
\begin{tabular*}{200pt}{@{\extracolsep{\fill}}ld{4.2}d{4.2}d{4.2}@{}}
\hline
${\bolds{l}}$ & \multicolumn{1}{c}{\textbf{1993--1995}} & \multicolumn{1}{c}{\textbf{1995--1998}} & \multicolumn{1}{c@{}}{\textbf{1998--2000}}\\
\hline
{1} & 544.8^{\ast} & 597.1 & 1141.8^{\ast} \\
{2} & 569.8 & 597.9 & 2040.8 \\
{3} & 634.7 & 590.5^{\ast} & \multicolumn{1}{c}{$2.9 \cdot e^{5}$} \\
{4} & 1230.2 & 592.0 & \multicolumn{1}{c}{$7.9 \cdot e^{6}$} \\
{5} & 2224.7 & 1352.6 & \multicolumn{1}{c}{$4.1 \cdot e^{7}$} \\
\hline
\end{tabular*}
\end{table}

%s5.3 #&#
\subsection{Assessment of overlap, balance, and unconfoundedness}
\label
{sec:assess}
\begin{figure}[b]

\includegraphics{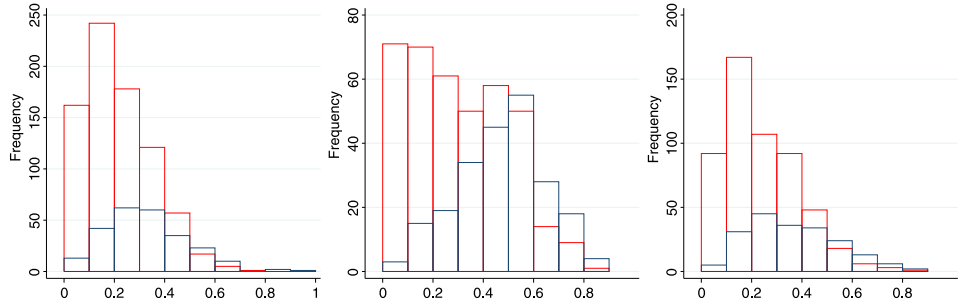}

\caption{Histograms of the estimated propensity score for the treated
(blue) and the untreated (red). The first is for the span 1993 to
1995, the second is for the span 1995 to 1998, and the third is for the
span 1998 to 2000.} \label{fig:hist_ps}
\end{figure}

We assess the overlap assumption by plotting the distribution of the
estimated propensity scores in the treatment and control groups and
visually inspecting the overlap. Figure~\ref{fig:hist_ps} presents the
histograms of the propensity scores estimated from the semiparametric
model \eqref{eq:ps}\vadjust{\goodbreak} for the treated and control groups, which shows a
satisfactory overlap in all three spans. Nevertheless, for the purpose
of further improving the overlap, we discard the very few units with
extreme values of the estimated propensity score: one unit for the span
1993--1995, and four units for the span 1995--1998.

%f2 #&#

\begin{figure}

\includegraphics{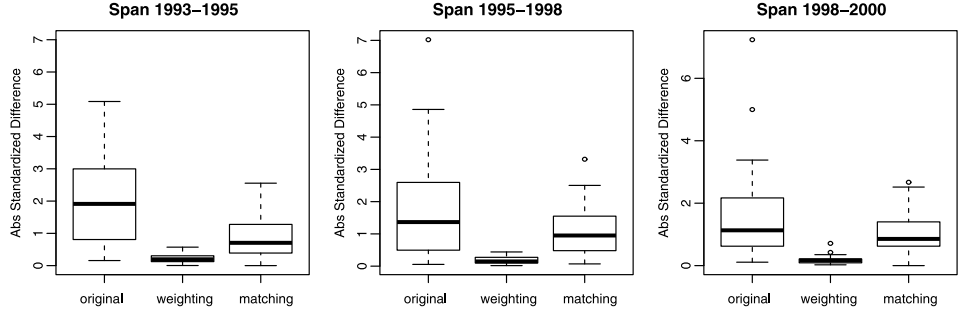}

\caption{Boxplots of the absolute standardized difference of all
covariates in the original, weighted, and matched data.} \label{fig:boxplot}
\end{figure}

We further check the balance of covariates based on the estimated
propensity score under each estimating method. In particular, we
measure covariate balance by the absolute standardized difference
(ASD), that is, the absolute difference in the means of the weighted
covariate between the treatment and control groups divided by the
square root of the sum of within group variances:
%
%e9 #&#
\begin{equation}
\operatorname{ASD} = {\biggl\llvert \frac{\sum_{i=1}^N X_i Z_i w_i}{\sum_{i=1}^N Z_i
w_i} - \frac{\sum_{i=1}^N X_i (1- Z_i) w_i}{\sum_{i=1}^N (1-Z_i)
w_i}\biggr
\rrvert } \Big/{\sqrt{s_{1}^2/N_1 +
s_{0}^2/N_0}},\label{eq:std_bias}
\end{equation}
where $N_z$ is the number of units and $s_z^2$ is the standard
deviation of the unweighted covariate in group $Z=z$ for $z=0,1$. For
the original data (used in the regression estimator $\hat{\tau
}_{\mathrm{reg}}$), $w_i=1$ for each unit and ASD is the standard two-sample
$t$-statistic; for the weighting-based estimators ($\hat{\tau}_{\mathrm{wt}}$ and
$\hat{\tau}_{\mathrm{dr}}$), $w_i$ are the ATT weights defined before; for the
matching-based estimator ($\hat{\tau}_{\mathrm{mix}}$), for each treated unit
$w_i$ equals 1 and for each control units $w_i$ equals the number of
that unit being sampled (can be larger than 1 in the case of matching
with replacement). The boxplots of the ASD for all covariates from
different methods are shown in Figure~\ref{fig:boxplot}. Weighting
leads to substantial improvement in the overall balance of all
covariates, with the largest ASD smaller than 1 (1.96 can be viewed as
a threshold of significant difference). This can be viewed as evidence
that the propensity score is well estimated.

For the mixed matching-regression estimator $\hat{\tau}_{\mathrm{mix}}$, the
number of matched units $m$ was set at 6, and the distance metric,
$\llVert \mathbf{x}\rrVert = ( \mathbf{x}^{\prime}
\mathbf
{S x} ) ^{1/2}$ where $\mathbf{S}$ is the diagonal matrix of the
inverses of the covariates variances, is adopted. From the boxplot we
can see matching ($m=6$) also improves balance, but significant
residual imbalance presents in several variables. Comparing covariate
balances and estimated effects obtained from matched samples with
different values of $m$ (2 to 6), we noticed a bias-variance trade-off
in $\hat{\tau}_{\mathrm{mix}}$: a larger number of matches ($m$) increases
residual imbalance in covariates, but at the same time decreases the
standard errors of the estimate. Because the regression step in $\hat
{\tau}_{\mathrm{mix}}$ can adjust for the residual imbalance, we choose $m=6$ in
the SHIW to reduce the standard errors.

%f3 #&#

The unconfoundedness assumption is generally untestable, and here we
adopt the approach of \citet{Cruetal08} to indirectly assess its
plausibility via a test based on quantifying the treatment effect on
the lagged outcome. The idea is that the lagged outcome, $Y_{\mathrm{lag}}$, can
be considered a proxy for $Y(0)$ and, given it is observed before the
treatment, it is unaffected by the treatment. Consequently, if the
average treatment effect on the lagged outcome is estimated to be zero
for all subpopulations defined by the rest of the pre-treatment
covariates, $\mathbf{V}=(\mathbf{X}\setminus Y_{\mathrm{lag}})$, then the
unconfoundedness assumption is plausible. The hypotheses are formalized as
\begin{eqnarray*}
&& H_{0}\dvtx\bE(Y_{\mathrm{lag},z=1}-Y_{\mathrm{lag},z=0}|\mathbf{V}=
\mathbf{v})=0 \qquad\forall\mathbf{v}\quad\mbox{vs.}
\\
&& H_{1}\dvtx\exists\mathbf{v }\mbox{ s.t. } \bE
(Y_{\mathrm{lag},z=1}-Y_{\mathrm{lag},z=0}|\mathbf{V}=\mathbf{v})\neq0,
\end{eqnarray*}
and can be tested using the aforementioned power series regression
approach to estimation for average treatment effects. In particular,
given the order term $k$, the series estimator of the regression
function $\mu_{z} ( \mathbf{v} ) $ of the lagged outcome
$Y_{\mathrm{lag}}$ under the treatment $z$ ($z=0,1$) is
\[
\hat{\mu}_{z} ( \mathbf{V} ) =\mathbf{D}_{k,z}\bigl(\mathbf
{D}%
_{k,z}^{\prime}\mathbf{D}_{k,z}
\bigr)^{-1}\mathbf{D}_{k,z}^{\prime
}\tilde {
\mathbf{Y}}_{\mathrm{lag},z}= \mathbf{D}_{k,z}\hat{\bolds{\xi}}_{k,z},
\]
where $\bD_{k,z}=\mathbf{P}_k(\tilde{\bV}_{z})$, with $\tilde{\bV}_{z},
\tilde{\mathbf{Y}}_{\mathrm{lag},z}$ being the matrices and the vector of all
the observed values of $\bV, \mathbf{Y}_{\mathrm{lag}}$ in group $z$ and
$\mathbf
{P}_k(\cdot)$ defined as in Section~\ref{sec:semi_spec}. \citet{Che07}
shows that if $k$ increases with the sample size $N$ (even if at a
lower rate), the test statistic $Q$ is the quadratic form and converges
to a chi-square distribution with $k$ degrees of freedom under the null
hypothesis:
\[
Q=(\hat{\bolds{\xi}}_{k,1}-\hat{\bolds{\xi}}_{k,0})^{\prime}
\hat{\mathbf{S}}_{k}^{-1}(\hat{\bolds{\xi}}_{k,1}-
\hat{\bolds {\xi }}_{k,0})\rightarrow \chi^{2}(k),
\]
where $\hat{\mathbf{S}}_{k}=\sum_{z}\mathbf{S}_{z}$ with $\mathbf
{S}_{z}$ being the estimated limiting variance of $\sqrt{N_{z}}
\hat{\bolds{\xi}}_{z,k}$. Therefore, implementation of the test is identical
to that of a parametric test for the equality restrictions
$\bolds{\xi}_{k,1}=\bolds{\xi}_{k,0}$ in the parametric setting.

%t6 #&#
\begin{table}[b]
\caption{Results from the tests to assess unconfoundedness}
\label{tab:unconfoundtest}
\begin{tabular*}{200pt}{@{\extracolsep{\fill}}ld{2.3}d{2.3}d{2.3}@{}}
\hline
& \multicolumn{1}{c}{\textbf{1993--1995}} & \multicolumn{1}{c}{\textbf{1995--1998}} & \multicolumn{1}{c@{}}{\textbf{1998--2000}}\\
\hline
${l}_{\min\ \mathrm{CV}}$ & 3 & 5 & 2 \\
${k}$ & 42 & 52 & 38 \\
${Q}$ & 52.7 & 43.9 & 36.9 \\
{$p$-value} & 0.124 & 0.780 & 0.522\\
\hline
\end{tabular*}
\end{table}

Table~\ref{tab:unconfoundtest} shows the values for $Q$ (along with
their respective $p$-values) under the null of unconfoundedness, where
$l$ is set to $l_{\min\ \mathrm{CV}}$, namely, the maximum power term in the
power series expansion of the nonparametric part $g(\bolds{\cdot})$ for which
the MSE is minimized. For 1995--1998 and 1998--2000, $l_{\min\ \mathrm{CV}}$ do not
coincide for the untreated, $l_{\min\ \mathrm{CV}}=5,1$, respectively, and the
treated units, $l_{\min\ \mathrm{CV}}=4,2$. For these two spans, $l_{\min\ \mathrm{CV}}$
has been posed at 5 and 2, respectively,\vadjust{\goodbreak} for the regressions $\mu
_{z} ( \mathbf{v} ) $ in order to undersmooth the
nonparametric specification. The $p$-values for the three periods (0.124,
0.780, 0.522 for period 1993--1995, 1995--1998, 1998--2000, resp.) suggest that
there is no difference in the lagged outcome between groups and,
consequently, the unconfoundedness assumption is deemed plausible.

%t7 #&#
\begin{table}
\tabcolsep=0pt
\caption{Estimated PATT in thousands of Italian Lira (standard errors
in parenthesis)}
\label{tab:result}
\begin{tabular*}{\textwidth}{@{\extracolsep{\fill}}lccccccccccccc@{}}
\hline
& & & \multicolumn{2}{c}{$\bolds{\hat{\tau}_{\mathrm{reg}}}$} && \multicolumn
{2}{c}{$\bolds{\hat
{\tau}_{\mathrm{wt}}}$}
& & \multicolumn{2}{c}{$\bolds{\hat{\tau}_{\mathrm{mix}}}$} && \multicolumn
{2}{c@{}}{$\bolds{\hat
{\tau}_{\mathrm{dr}}}$} \\[-4pt]
& & & \multicolumn{2}{c}{\hrulefill} && \multicolumn
{2}{c}{\hrulefill}
& & \multicolumn{2}{c}{\hrulefill} && \multicolumn
{2}{c@{}}{\hrulefill} \\
\textbf{Span} & \textbf{AOT} & & \textbf{PATT} & $\frac{\mbox{\textbf{PATT}}}{\mbox{\textbf{AOT}}}$ & & \textbf{PATT} &
$\frac{\mbox{\textbf{PATT}}}{\mbox{\textbf{AOT}}}$ & & \textbf{PATT} & $\frac{\mbox{\textbf{PATT}}}{\mbox{\textbf{AOT}}}$
& & \textbf{PATT} & $\frac{\mbox{\textbf{PATT}}}{\mbox{\textbf{AOT}}}$ \\
\hline
{1993}--{1995} & 2092.9 & & 90.2 & 0.043 & & 102.3 & 0.049 & & 100.6 &
0.048 & & 97.2 & 0.046 \\
& & & (41.8) & & & (47.1) & & & (50.4) & & & (42.7) & \\
{1995}--{1998} & 2027.6 & & 199.1 & 0.098 & & 160.7 & 0.079 & & 208.7 &
0.103 & & 202.2 & 0.100 \\
& & & (87.6) & & & (73.4) & & & (69.8) & & & (93.2) & \\
{1998}--{2000} & 2116.4 & & 148.1 & 0.069 & & 137.7 & 0.065 & & 122.8 &
0.058 & & 142.1 & 0.067 \\
& & & (68.5) & & & (73.1) & & & (60.7) & & & (70.5) & \\
\hline
\end{tabular*}\vspace*{-5pt}
\end{table}

%s5.4 #&#
\subsection{Results}
We obtain results from the regression estimator $\hat{\tau}_{\mathrm{reg}}$, the
propensity score weighting estimator $\hat{\tau}_{\mathrm{wt}}$, and the DR
estimator $\hat{\tau}_{\mathrm{dr}}$ by using common routines for linear and
logistic regressions in STATA and GAUSS. Point estimates and standard
errors from the mixed matching-regression estimator $\hat{\tau}_{\mathrm{mix}}$
have been obtained by the STATA program by \citet{Abaetal03}.

Table~\ref{tab:result} reports the effects of debit cards on household
monthly consumption estimated from the four estimators. The ratios
between each estimated PATT and the Average Outcome for the Treated
(AOT) are also reported. Positive and statistically significant
estimates of the PATT are consistently obtained across all estimators
and all spans. For the span of 1993--1995, the increase in the monthly
consumption for the household with one debit card ranges from 4.3\% to
4.9\% (90.2 to 100.6 thousands Italian Liras) across the four
estimators; for the span of 1998--2000, the increase ranges from 5.8\% to
6.9\% (122.8 to 148.1 thousands Liras). The period 1995--1998 emerges with
particular high estimated PATT, ranging from 7.9\% (160.7 thousands
Liras) to 10.3\% (208.7 thousands Liras) of the household monthly
consumption. This can be explained by the fact that the period was
observed one year longer (the planned survey for 1997 was delayed one
year and shifted to 1998); therefore, the use of debit cards could have
benefited from the longer span to more strongly affect consumers'
behavior. Note that in the span of 1995--1998, the weighting estimate is
significantly different from both the regression and the DR estimates,
suggesting a potential misspecification of the propensity score.

Overall, these results support current psychological theories about the
effects of debit cards on spending. As debit cards do not allow the
consumer to incorporate an additional\vadjust{\goodbreak} long-term source of funds, our
analysis largely eliminates the potential confounded effect of an
intertemporal reallocation of wealth [\citet{SomChe02}].
Therefore, the significant estimated effects on spending can be
ascribed only to psychological reasons, in particular, those pertaining
to the aforementioned theories regarding the rehearsal of the price
paid [\citet{Som01}] and regarding the accessibility of financial
resources [Morewedge, Hotzman and
Epley (\citeyear{MorHotEpl07})]. Both theories state that payments
by debit cards enlarge the perceived amount of financial resources
available for consumption compared to pay cash. This is due, in one
case, to an impact on the memory of past expenses and, in the other
case, to the cognitive accessibility to a larger financial resource
like the savings account. Our findings can be explained by the
microeconomic theory of consumer choice in that the perception of
larger disposable financial resources implies less budget constraints.
This enlarges the set of affordable bundles and increases the ordinary
demand because the most preferred affordable bundle, that is, the
rational consumer's choice, will be composed by a larger quantity of
goods for rational utility functions. This is essentially an income
effect; however, if the occasions to pay by debit cards differ by
categories of goods, also a substitution effect will be in act. The
latter could be studied by evaluating the effect on spending for
different categories of goods, for example, for food versus nonfood consumption.

%s6 #&#
\section{Conclusion} \label{sec:conclusion}
Motivated by recent findings in the field of consumer science, we
conduct a population-based study based on the Italy SHIW data to
evaluate the causal effect of debit cards on household consumption.
Within the RCM, we adopt several power series-based semiparametric
approaches to estimate the PATT. The key assumptions, overlap and
unconfoundedness, are systematically assessed and validated. Our
analysis suggests that possessing debit cards significantly increases
monthly household spending in Italy, consistent with and complementary
to the findings from several small randomized experiments in psychology
and consumer science.

One limitation of the study is that only short-time effects of the
considered payment instrument have been here evaluated. In fact, the
SHIW data do not provide information about the moment a treated
household has acquired its debit cards. We only know it has happened
during the two, or three, years of the considered span, so that we have
likely estimated one to one-and-a-half years long effects. A desirable
extension of this work may be to apply the same causal methods to
suitable data sets that allow for enlarging the extent of the temporal
effects of debit cards. Another limitation is that, due to data
availability, this study focuses on household rather than individual
spending. This problem is partially mitigated by restricting the
analysis to households with one and only one debit card. Nevertheless,
an analysis on singleton households or households formed only by a
couple or population-based data with individual information would
provide more information.\vadjust{\goodbreak}

We have focused on the PATT estimand; other estimands may be of
interest depending on the study goal. For example, if the goal were to
plan a marketing policy aimed to increase spending by stimulating the
use of noncash payment instruments, then the relevant causal effect
should be on untreated units. Consequently, the estimators considered
here need to be modified accordingly.

\begin{appendix}\label{app}
\section*{Appendix: Proof of the ``double robustness'' property of the
estimator
\texorpdfstring{$\hat{\tau}_{\mathrm{dr}}$}{hattaudr}}
It is trivial to prove the first term of (\ref
{eq:est_dr}), $\sum_{i=1}^{N}Y_{i}Z_i/\sum_{i=1}^{N}Z_i$, converges to
$\bE\{Y(1) | Z=1\}$.

For the second term, first suppose the outcome model $\mu_{0}(\bX)$ is
correctly specified but the propensity score $e(\bX)$ is misspecified,
so that $\hat{\mu}_{0}(\bX)\rightarrow\bE(Y(0) | \bX)$, $\hat
{e}(\bX
)\nrightarrow\Pr(Z=1 | \bX)$. Then we have
\begin{eqnarray}
\label{eq:proof} &&\frac{1}{\sum_{i=1}^{N}Z_i}\sum_{i=1}^{N}
\frac{Y_{i}(1-Z_i)\hat
{e}(\bX
_{i})+\hat{\mu}_{0}(\bX_{i})(Z_i-\hat{e}(\bX_{i}))}{1-\hat{e}(\bX_{i})}
\nonumber
\\
&&\qquad=\frac{1}{\sum_{i=1}^{N}Z_i} \biggl\{ \sum_{i:Z_i=1}\hat{\mu
}_{0}(\bX _{i})+\sum_{i:Z_i=0}
\frac{Y_{i} \hat{e}(\bX_{i})-\hat{\mu
}_{0}(\bX_{i})
\hat{e}(\bX_{i})}{1-\hat{e}(\bX_{i})} \biggr\}
\nonumber
\\[-8pt]
\\[-8pt]
\nonumber
&&\qquad=\frac{1}{\sum_{i=1}^{N}Z_i}\sum_{i:Z_i=1}\hat{
\mu}_{0}(\bX _{i})\\
&&\qquad\quad{}+\frac
{\sum_{i=1}^{N}(1-Z_i)}{\sum_{i=1}^{N}Z_i}\frac{1}{\sum_{i=1}^{N}(1-Z_i)} \sum
_{i:Z_i=0} \biggl\{ \frac{Y_{i} \hat{e}(\bX_{i})}{1-\hat{e}(\bX_{i})}- %%
\frac{\hat{\mu}_{0}(\bX_{i}) \hat{e}(\bX_{i})}{1-\hat{e}(\bX
_{i})} \biggr\}.
\nonumber
\end{eqnarray}
It is straightforward to prove, given the law of large numbers and the
consistency of $\hat{\mu}_{0}(\bX)$, that
$\sum_{i:Z_i=1}\hat{\mu}_{0}(\bX_{i}) /\sum_{i=1}^{N}Z_i$ converges
to $\bE_{\bX} \{ \bE ( Y(0)|\bX ) |\break  Z=1 \}$, and
$\sum_{i:Z_i=0}\frac{Y_{i} \hat{e}(\bX_{i})}{1-\hat{e}(\bX
_{i})}
/ {\sum_{i=1}^{N}(1-Z_i)}$ and $\sum_{i:Z_i=0}\frac{\hat{\mu
}_{0}(\bX
_{i}) \hat{e}(\bX_{i})}{1-\hat{e}(\bX_{i})} /\break{\sum_{i=1}^{N}(1-Z_i)}$ converge to the same quantity $\bE_{\bX} \{
\bE
 (Y(0) | \bX ) \frac{\hat{e}(\bX)}{1-\hat{e}(\bX)}
 |
Z=0 \}$. Consequently, equation \eqref{eq:proof} converges to (the
subscript $i$ is dropped to simplify the notation)
\begin{eqnarray*}
\bE_{\bX} \bigl\{ \bE \bigl( Y(0) | \bX \bigr) | Z=1 \bigr\} &=&\int
_{\mathbf{x}}f (\mathbf{x}| Z=1 ) \bE \bigl\{ Y(0) | \bX=\mathbf{x}
\bigr\}
\\
&=&\int_{\mathbf{x}}f (\mathbf{x}| Z=1 ) \bE \bigl\{ Y(0) | \bX=
\mathbf{x} , Z=1 \bigr\}
\\
&=&\int_{\mathbf{x}}f ( \mathbf{x}| Z=1 ) \int_{y(0)}y(0)
f \bigl( y(0) | \bX =\mathbf{x},Z=1 \bigr)
\\
%&=&\int_{y(0)}y(0)\int_{\bx}f\left( y(0) | \bX=\bx,Z=1\right) f\left(
%&=&\int_{y(0)}y(0)\int_{\bx}f\left( y(0),\bX=\bx| Z=1\right) \\
&=&\int_{y(0)}y(0) f \bigl(
y(0) | Z=1 \bigr)
\\
&=&\bE \bigl\{Y(0) | Z=1 \bigr\}.
\end{eqnarray*}
Above, the second equation is due to the unconfoundedness assumption.

Alternatively, suppose $e(\bX)$ is correctly specified while $\mu
_{0}(\bX)$ is misspecified, so that $\hat{e}(\bX)\rightarrow\Pr
(Z=1 | \bX)$, $\hat{\mu}_{0}(\bX)\nrightarrow\bE(Y(0) | \bX)$. Again,
it is easy to prove, given the law of large numbers and the
consistency of $\hat{e}(\bX_{i})$, that the quantities $\sum_{i:Z_i=1}\hat{\mu}_{0}(\bX_{i})/{\sum_{i=1}^{N}Z_i}$ and $\frac
{\sum_{i=1}^{N}(1-Z_i)}{\sum_{i=1}^{N}Z_i}\frac{1}{\sum_{i=1}^{N}(1-Z_i)}\times\break \sum_{i:Z_i=0} \{
\frac{\hat{\mu}_{0}(\bX_{i}) \hat{e}(\bX_{i})}{1-\hat{e}(\bX
_{i})}
\} $
converge to the same quantity $\bE_{\bX} \{ \hat{\mu}_{0}(\bX
) | Z=0 \} $, and $\frac{\sum_{i=1}^{N}(1-Z_i)}{\sum_{i=1}^{N}Z_i}
\frac{1}{\sum_{i=1}^{N}(1-Z_i)} \sum_{i:Z_i=0} \{
\frac{Y_{i} \hat{e}(\bX_{i})}{1-\hat{e}(\bX_{i})} \} $ converges
to $\frac{\Pr(Z=0)}{\Pr(Z=1)}\bE_{\bX} \{ \frac{\Pr(Z=1 |
\bX
)}{\Pr(Z=0 | \bX)}\times\break 
\bE ( Y(0) | \bX )  | Z=0 \} $.
Consequently, equation \eqref{eq:proof} converges to
\begin{eqnarray*}
&&\frac{\Pr(Z=0)}{\Pr(Z=1)}\bE_{\bX} \biggl\{ \frac{\Pr(Z=1 | \bX
=\mathbf{x}
)}{\Pr(Z=0 | \bX=\mathbf{x})} \bE \bigl(
Y(0) | \bX \bigr) | Z=0 \biggr\}
\\
&&\qquad=\frac{\Pr(Z=0)}{\Pr(Z=1)}\int_{\mathbf{x}}f (\mathbf{x}| Z=0 )
\frac{\Pr
(Z=1 | \bX=\mathbf{x})}{\Pr(Z=0 | \bX=\mathbf{x})}\int_{y(0)}y(0) f \bigl( y(0) | \bX =
\mathbf{x} \bigr)
\\
%&=&\frac{\Pr(Z=0)}{\Pr(Z=1)}\int_{\bx}f\left( \bx| Z=0\right) \frac{
%y(0) | \bX=\bx, Z=1\right) \\
&&\qquad=\int_{\mathbf{x}} f (
\mathbf{x}| Z=0 )\frac
{f(\mathbf{x}| Z=1)}{f(\mathbf{x}| Z=0)} \int_{y(0)}y(0) f \bigl(y(0)
| \bX=\mathbf{x}, Z=1 \bigr)
\\
&&\qquad=\int_{y(0)}y(0)\int_{\mathbf{x}} f(\mathbf{x}|
Z=1) f \bigl( y(0) | \bX=\mathbf{x} , Z=1 \bigr)
\\
%&=&\int_{y(0)}y(0)\int_{\bx} f\left( y(0),\bx| Z=1\right) \\
&&\qquad=\int_{y(0)}y(0) f \bigl( y(0) | Z=1 \bigr)
 =\bE \bigl\{Y(0) | Z=1 \bigr\}.
\end{eqnarray*}

The same arguments apply to the case of both $\mu_{0}(\bX)$ and
$e(\bX
)$ correctly
specified. Then, $\hat{\tau}_{\mathrm{dr}}$ converges to $\bE\{Y(1) | Z=1\}
-\bE\{
Y(0) | Z=1\}=\tau^{\PATT}$ when the outcome model $\mu_{0}(\bX)$ and/or
the propensity score model $e(\bX)$ are correctly specified. \qed
\end{appendix}
\section*{Acknowledgments}
We thank the Associate Editor and two anonymous reviewers for
constructive comments that help improve the clarity and exposition of
the article, and Quanli Wang for computing support. The content is
solely the responsibility of the authors and does not necessarily
represent the official views of Bank of Italy, SAMSI, or NSF.

% imsref loaded by akundreckaite, 2014-10-14 15:44:43
% imsref loaded by akundreckaite, 2014-10-15 08:01:09
%
% imsref loaded by akundreckaite, 2014-10-15 08:39:32
% imsref loaded by akundreckaite, 2014-10-15 08:40:36

%


%suskaldyti doi

\printaddresses
\end{document}